\documentclass[10pt]{iopart}
\usepackage{iopams}
\usepackage{cite}
\usepackage[normalem]{ulem}
\usepackage[utf8]{inputenc}
\usepackage{xspace}
\usepackage{graphicx}
\usepackage{epstopdf}
\usepackage{color,xcolor}
\usepackage{upgreek}
\usepackage{nicefrac}
\usepackage{siunitx}

\DeclareSIUnit\atper{{at.\,\percent}}
\usepackage{url}
\usepackage[colorlinks=true]{hyperref}

\newcommand{\EF}{E_\text{F}}         
\newcommand{\muB}{\mu_\text{B}}      
\newcommand{\Tc}{T_\text{C}}          

\newcommand{\VO}{V$_\text{O}$}
\newcommand{\ag}{\ensuremath{A_{1\text{g}}}}
\newcommand{\bag}{\ensuremath{B_{1\text{g}}}}
\newcommand{\bbg}{\ensuremath{B_{2\text{g}}}}
\newcommand{\eg}{\ensuremath{e_{\text{g}}}}
\newcommand{\due}{e_\text{g}^{\uparrow}}

\newcommand{\dut}{t_\text{2g}^{\uparrow}}

\newcommand{\tg}{\ensuremath{t_{2\text{g}}}}
\newcommand{\dxy}{\ensuremath{d_{xy}}}
\newcommand{\dxz}{\ensuremath{d_{xz}}}
\newcommand{\dyz}{\ensuremath{d_{yz}}}
\newcommand{\dzz}{\ensuremath{d_{3z^2}}}
\newcommand{\dxxyy}{\ensuremath{d_{x^2-y^2}}}

\newcommand{\dd}{\ensuremath{d^{\downarrow}}}

\newcommand{\affJKU}{Institute for Theoretical Physics, Johannes Kepler University Linz, Altenberger Straße 69, 4040 Linz, Austria}
\newcommand{\affmlu}{Institute of Physics, Martin Luther University Halle-Wittenberg, Von-Seckendorff-Platz 1, 06120 Halle, Germany}

\newcommand{\affmpihalle}{Max Planck Institute of Microstructure Physics, Weinberg 2, 06120 Halle, Germany}
\newcommand{\affUniTurku}{Department of Physics and Astronomy, University of Turku, FIN-20014 Turku, Finland}
\newcommand{\affMatSurf}{Turku University Centre for Materials and Surfaces (MatSurf), Turku, Finland}
\newcommand{\affKiev}{G. V. Kurdyumov Institute for Metal Physics of the N.A.S. of Ukraine, 36 Vernadsky Street, 03142 Kiev, Ukraine}
\newcommand{\affPoland}{Faculty of Mathematics and Informatics, University of Bialystok, K. Ciolkowskiego 1M, PL-15-245 Bialystok, Poland}

\usepackage{dcolumn}
\newcolumntype{.}{D{.}{.}{4}}
\newcolumntype{;}{D{.}{.}{4}}
\usepackage{pdfpages}
\newcommand{\UFe}{U^\text{Fe}_\text{eff}}

\newcommand{\Feval}[1]{Fe$^{#1 +}$}
\newcommand{\Moval}[1]{Mo$^{#1 +}$}
\newcommand{\sikelvin}[1]{\SI{#1}{\kelvin}\xspace}
\newcommand{\sieV}[1]{\SI{#1}{\eV}\xspace}

\newcommand{\simB}[1]{#1\,\muB\xspace}
\newcommand{\Ledge}[1][]{{L$_{#1}$}\xspace}
\newcommand{\Kedge}[1][]{{K$_{#1}$}\xspace}
\newcommand{\Medge}[1][]{{M$_{#1}$}\xspace}
\begin{document}
\title{Variation of magnetic properties of Sr$_2$FeMoO$_6$ due to oxygen vacancies}

\author{Martin Hoffmann$^{1}$, Victor N. Antonov,$^{2,3}$ Lev V. Bekenov,$^{2,4}$, 
        Kalevi Kokko,$^{5,6}$, Wolfram Hergert,$^{7}$, and Arthur Ernst$^{1,4}$}
\address{$^1$\affJKU}
\address{$^2$\affKiev}
\address{$^3$\affPoland}
\address{$^4$\affmpihalle}
\address{$^5$\affUniTurku}
\address{$^6$\affMatSurf}
\address{$^7$\affmlu}
\ead{martin.hoffmann@jku.at}

\date{\today}

\begin{abstract}
Oxygen vacancies can be of utmost importance
for improving or deteriorating physical properties of oxide materials. Here, we studied from
first-principles the electronic and magnetic properties of oxygen vacancies in the double
perovskite Sr$_2$FeMoO$_6$ (SFMO). We show that oxygen vacancies can increase the
Curie temperature in SFMO, although the total magnetic moment is reduced at
the same time. We found also that the experimentally observed valence change of
the Fe ions from $3+$ to $2+$ in the x-ray magnetic circular dichroism (XMCD) measurements
is better explained by oxygen vacancies than by the assumed mixed valence state.
The agreement of the calculated x-ray absorption spectra and XMCD results with
experimental data is considerably improved by inclusion of oxygen vacancies.
\end{abstract}

\maketitle

\ioptwocol
\section{Introduction}
\label{sec:int}

Double perovskites $A_2BB^{\prime}$O$_6$ ($A$ = alkaline earth or rare
earth metal atoms and $BB^{\prime}$ = heterovalent transition metal atoms such as $B$
= Fe, Cr, Mn, Co, Ni; $B^{\prime}$ = Mo, Re, W) often demonstrate
intrinsically complex magnetic structures and a wide variety of
physical properties (see \cite{Serrate2007jpcm} for a
review article on these materials).
In order to understand the complex properties,
experiments on SFMO included numerous methods like photoemission spectroscopy (PES) 
\cite{Saitoh2002PRB,Kang2002prb,Kim2003jkps}, M\"ossbauer spectroscopy \cite{Sarma2000ssc, Linden2000apl}, 
neutron scattering \cite{Kapusta2002jmmm,Garca-Landa1999,Moritomo2000jpsj},
x-ray absorption spectroscopy (XAS), and x-ray magnetic circular
dichroism (XMCD) measurements 
\cite{Besse2002epl,Kuepper2004pssa,Koide2014jpcser}.
Since the XAS and XMCD spectroscopy is very
sensitive to the electronic structure and the local environment, 
comparing calculated spectra to the available
experimental results can provide important information about
the chemical composition, the ionic valency, and the degree of
electronic correlations in the system. 
Thus, we simulated 
the  XAS and XMCD spectra with our method and found 
a better agreement than previous density
functional calculations \cite{Kanchana2007prb}.
Our results indicate as well a mixed valency of the Fe ion (Fe$^{2+}$ or Fe$^{3+}$)
observed by experiments \cite{Linden2000apl, Kapusta2002jmmm, Besse2002epl, Kuepper2004pssa},
but we can also conclude from our study that the mixed valency is mainly caused by oxygen vacancies.

In addition, several experimental and theoretical studies have demonstrated that
the double perovskite system Sr$_2$FeMoO$_6$ (SFMO) and other related materials
exhibit a ferrimagnetic (FiM) half-metallic ground state with a high Curie
temperature of \SIrange{324}{420}{\kelvin} \cite{Kato2004prb,Rubi2006jpcs, Paturi2011, Saloaro2015}. The physical
origin of the magnetoresistance in SFMO is half-metallicity
\cite{Kobayashi1998n}, i.e., the material is an insulator in one of the
spin channels, but a metal in the other. This leads to a complete spin
polarization at the Fermi level, which immediately suggests their
application as a source of spin polarized charge carriers in
spintronic devices. Therein, SFMO will be used mainly as 
a thin film, and many attempts were made to grown high quality 
films \cite{Venimadhav2004jmmm, Westerburg2000prb, Kumar2010pb, Paturi2011, Saloaro2015}.
But all these films yield a reduced Curie temperature, which is 
up to \SI{80}{\kelvin} smaller than for bulk samples
\cite{Paturi2011, Saloaro2015}. Even for bulk SFMO
the theoretical magnetic moment of $\simB{4}$ was rarely
experimentally observed \cite{Garca-Landa1999}. Such variations were attributed to lattice 
defects like grain boundaries or point defects. In particular,
swapping of Fe and Mo ions, antisite disorder (ASD), 
\cite{Ogale1999apl, Solovyev2002prb, Colis2005jap_10.1063/1.1997286,
  Stoeffler2005jpcm, Munoz-Garcia2011, 
  Erten2011prl, Reyes2016jpcc} 
and oxygen vacancies (V$_\text{O}$)
\cite{Stoeffler2005jpcm, Munoz-Garcia2011, Kircheisen2012,
  Wu2014ssc}
were shown to reduce
the magnetization of SFMO.

Nevertheless, no study addressed 
explicitly the variation of the Curie 
temperature, $\Tc$, with oxygen vacancies (\VO).
We deploy therefore the Korringa-Kohn-Rostoker Green's function (KKR-GF) method 
with the coherent potential approximation (CPA).
We calculated magnetic coupling constants
using the magnetic
force theorem\cite{Liechtenstein1987jmmm} and used it with a 
classical Heisenberg model in order to
calculate the Curie temperature similarly done as in Refs. 
\cite{Fischer2009prb,Hoffmann2015prb_own}.
A similar approach handling ASD from first-principles 
was beyond the scope of this work, since the distribution of 
those antisite defects has to be studied carefully. A simple random
distribution might be ruled out by Ref. \cite{Meneghini2009prl}
showing a tendency for antiphase patches, which makes the study 
of the magnetic coupling of antisite defects a work on its own.

In the following section, we describe the lattice structure of SFMO
used in our calculations and give details about the calculation techniques.
In Sec.~\ref{sec:XASandXMCD}, we present the XAS and XMCD calculations
for the SFMO compound and demonstrate that oxygen vacancies have to 
be present in SFMO samples. 
The influence of oxygen vacancies on the electronic structure 
and magnetic properties is then
discussed in Sec.~\ref{sec:oxygen_vacancies}.

\section{Numerical Details}
\label{sec:structure_details}

\begin{figure}
  \includegraphics[width=236.68pt]{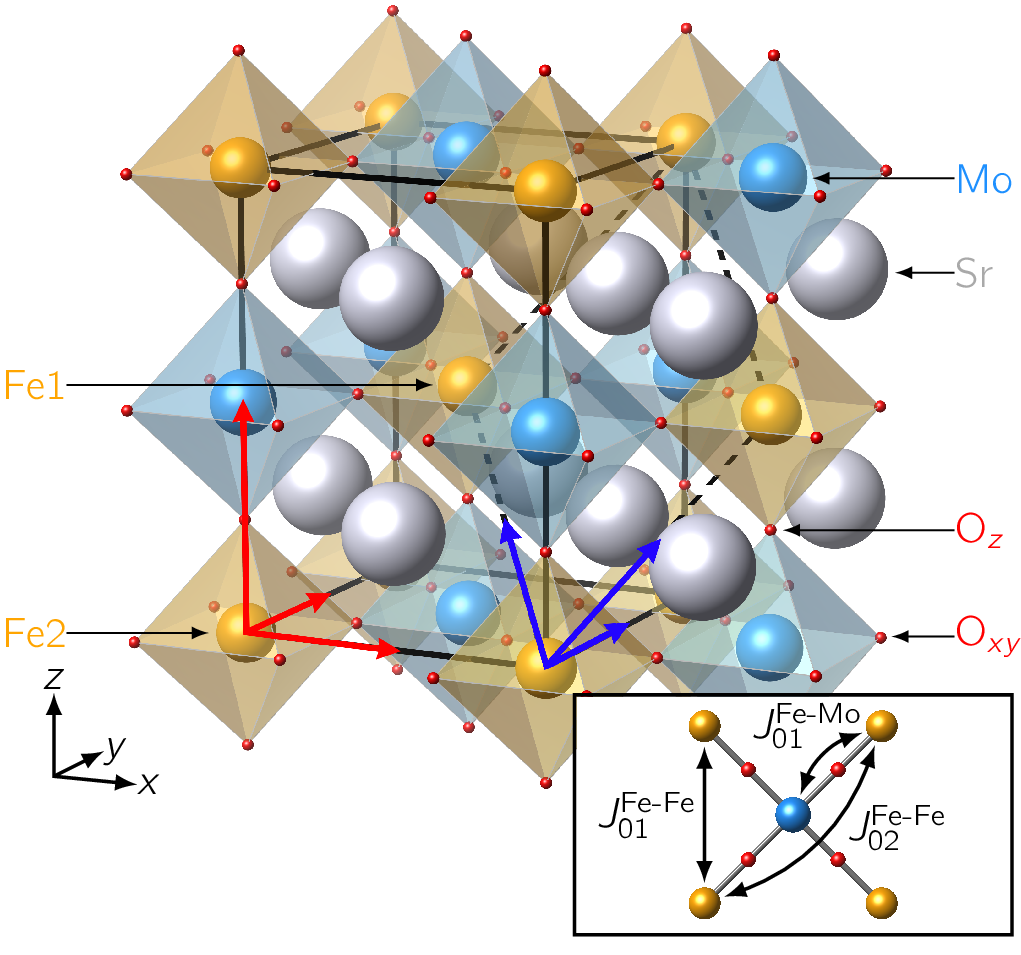}
  \caption{(Color online) The double perovskite structure of SFMO
    ($a=b=5.573$ and $c=7.902$ of Ref. \cite{Burzo2011}).
    The colored polyhedra visualize the octahedral surroundings of the
    Fe and Mo atoms (orange and blue).  
    Following from the tetragonal
    symmetry, two different oxygen positions appear (marked with
    O$_{xy}$ and O$_{z}$). The tetragonal supercell is shown by the
    black solid lines and the red arrows.  It contains two functional
    units with two Fe sites (Fe1, Fe2).  The black dashed lines and
    the blue arrows indicate the primitive unit cell. The inset shows
    the top view displaying the three most important magnetic coupling pairs
    between Fe-Fe and Fe-Mo.
    The figure was prepared with VESTA \cite{Momma2011jac}.
    }
  \label{fig:SFMO_structure}
\end{figure}

For our calculations of SFMO, we adopted the experimentally found double
perovskite structure, where the oxygen atoms provide an octahedral environment
around the Fe and Mo sites. The FeO$_6$ and
MoO$_6$ octahedra alternate along the three cubic axes, while the Sr atoms
occupy the hollow site formed by the corners of the FeO$_6$ and MoO$_6$
octahedra at the body-centered positions 
(\fref{fig:SFMO_structure}).

SFMO was found to be cubic ($Fm\bar{3}m$) in the paramagnetic phase,
but changes into a tetragonal-type structure below a critical
temperature \cite{Moritomo2000jpsj}. 
The Sr atoms occupy the 4$d$ Wyckoff positions 
(0,\,$\nicefrac{1}{2}$,\,$\nicefrac{1}{4}$).  The Fe
atoms occupy the 2$a$ Wyckoff pos.  (0,\,0,\,0) and Mo the 4$d$ Wyckoff
pos. (0,\,0,\,$\nicefrac{1}{2}$).  
There are two types of oxygen atoms with O$_z$ at the 4$e$
Wyckoff pos. (0,\,0,\,$z$) and O$_{xy}$ at the 8$h$ Wyckoff pos. 
($x$,\,$y$,\,0) (\fref{fig:SFMO_structure}).
Their positions is not definite and varies between different studies.
They occupy either the site exactly between Mo and Fe,
which gives the body centered tetragonal structure type ($I4/mmm$;
No. 139 \cite{Park2004pb, Burzo2011}) or the oxygen octahedra are
slightly distorted ($I4/m$, No. 87  \cite{Nakamura2003,Azad2006jssc}).
We used for
simplicity the more symmetric body centered tetragonal structure type
and the lattice constant an internal parameters from \cite{Burzo2011} as input for
our study.  

We note as well that the slight tetragonal distortion
is lifting the degeneration of the $d$ states.
The crystal field at the Fe (Mo) site (now $D_\text{4h}$ point
symmetry) splits the Fe (Mo) 3$d$ (4$d$) orbitals into three singlets
$\ag$ ($\dzz$), $\bag$ ($\dxxyy$), and $\bbg$ ($\dxy$) and a doublet
$\eg$ ($\dyz$ and $\dxz$).  Since the deviation
from the ideal $c/a$ ratio is only small, the states $\dxy$ and $\dzz$, as well as
the states $\dxxyy$, $\dyz$ and $\dxz$, form two groups of states,
which showed an almost similar density of states (DOS). Those are considered as $\eg$ and $\tg$
in accordance with other publications.

For the microscopic understanding of the SFMO compound, we combined
the theoretical results of three computational methods, namely the
multiple scattering KKR-Green's function method HUTSEPOT
\cite{Luders2001jpcm,Luders2005prb},
the spin-polarized
fully relativistic linear-muffin-tin-orbital (SPR-LMTO)
method \cite{Anderson1975prb,Nemoshkalenko1983pssb,Antonov1995jmmm} and
the Vienna \textit{ab initio} simulation package (VASP)
\cite{Blochl1994prb-1,Kresse1999prb}. The main 
investigation of the electronic and magnetic properties
were conducted with HUTSEPOT, whereas the XAS and XMCD spectra were 
calculated within the SPR-LMTO, and necessary structure relaxations
were achieved with VASP. All relevant input parameters and the 
discussion about the correct treatment of the electronic structure of SFMO
are provided in the Supporting Information (SI).

\begin{figure}
  \includegraphics[width=236.68pt]{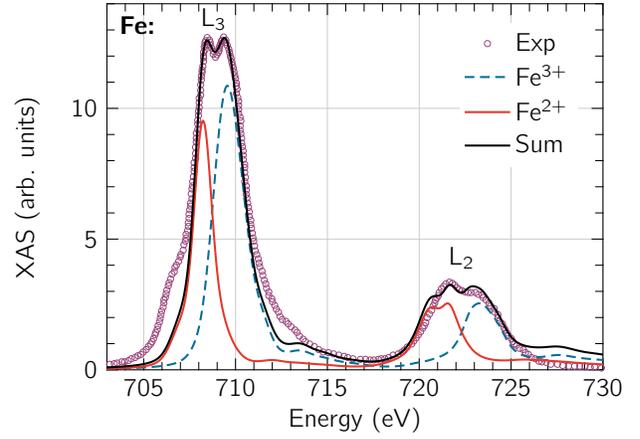}
  \caption{(Color online) The x-ray absorption spectra
    (Ref.~\cite{Besse2002epl}, open circles) at Fe \Ledge[2,3]
    edges as average of left and right circularly polarized light in
    SFMO measured at \sikelvin{10} with \SI{5}{\tesla} magnetic field
    compared with the theoretically calculated ones for Fe$^{3+}$
    (dashed blue line) and Fe$^{2+}$ (solid red line).
    The linear combination of \SI{60}{\percent} Fe$^{3+}$ and 
    \SI{40}{\percent} Fe$^{2+}$ denoted as Sum (solid black line)
    resembles best the reference curve. }
  \label{Fe_XAS}
\end{figure}

\section{X-ray absorption and XMCD spectra}
\label{sec:XASandXMCD}

The valency of the Fe ions is not finally resolved. Experimental studies 
find a mixed valency state of Fe$^{2+}$ and Fe$^{3+}$
\cite{Linden2000apl, Kapusta2002jmmm,Kuepper2004pssa,Besse2002epl}
but theoretical calculations result a ground state with Fe$^{3+}$
\cite{Solovyev2002prb,Szotek2003prb,Munoz-Garcia2011}. In order to solve
this discrepancy, the x-ray absorption and XMCD spectra
at the \Ledge[2,3] absorption edges can be used 
in a complex transition metal ionic compound such as SFMO
as fingerprints of the ground state. 

We used a supercell approximation with a cell
of two functional units of SFMO (see red arrows in
\fref{fig:SFMO_structure}).  This supercell was relaxed with
VASP including a single oxygen vacancy close to Fe2. The
relaxation with VASP for a single oxygen vacancy led to 
a distance of \SI{1.9722}{\angstrom} between Fe2 and the vacancy --
an increase by \SI{1.5}{\percent}.
The correspondent distance
between the Fe1 ion and the oxygen vacancy was 
reduced by \SI{0.5}{\percent} to \SI{4.4054}{\angstrom}.
In order to crosscheck effects of ASD on the calculated spectra,
we included also one antisite defect into the supercell.
We found, similarly as in
\cite{Munoz-Garcia2011}, no significant internal relaxations.

\begin{figure}
  \includegraphics[width=236.68pt]{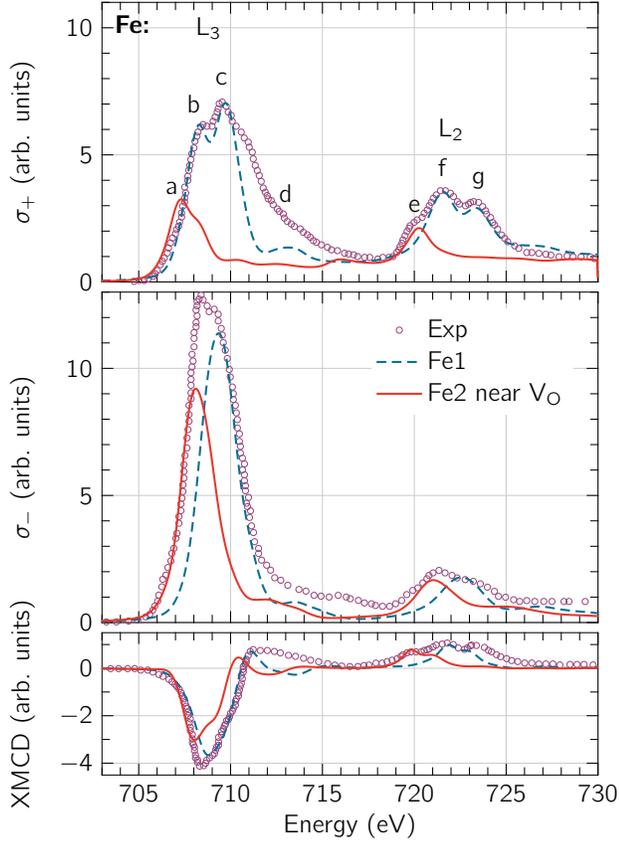}
  \caption{(Color online) The x-ray absorption spectra
    (Ref.~\cite{Besse2002epl}, open circles) at Fe \Ledge[2,3]
    edges in SFMO measured with left ($\sigma^+$, top panel) and right
    circularly polarized light ($\sigma^-$, middle panel) measured at
    \sikelvin{10} with \SI{5}{\tesla} magnetic field and XMCD
    experimental spectrum (Ref.~\cite{Besse2002epl}, lower
    panel) compared with the theoretically calculated spectra with an oxygen
    vacancy far away (dashed blue line) and near to an   
    Fe ion (solid red line). }
  \label{Fe_XMCD}
\end{figure}

\begin{figure}
  \includegraphics[width=236.68pt]{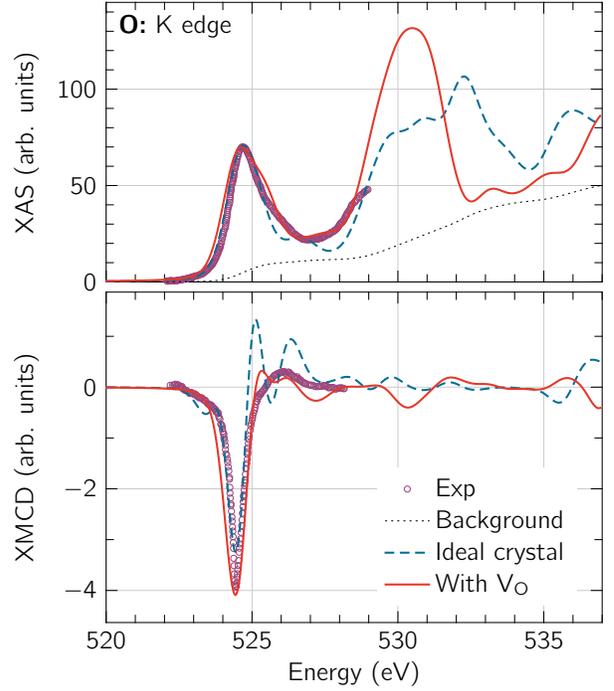}
  \caption{(Color online) The experimental x-ray absorption spectra
    (Ref.~\cite{Koide2014jpcser}, open circles) at O \Kedge edge
    (top panel) in SFMO and experimental XMCD spectra (lower panel)
    measured at $T=\sikelvin{20}$ with $B=\SI{1.1}{\tesla}$ compared
    with the theoretical simulations carried out for the ideal crystal
    structure (solid red line) and with an oxygen vacancy (dashed blue
    line).}
  \label{O_K}
\end{figure}

The experimentally measured Fe \Ledge[2,3] absorption
spectrum of the SFMO single crystal (Ref.~\cite{Besse2002epl},
average of left and right circularly polarized light) displays at both
absorption edges a weak lower-energy shoulder together with a doublet
structure at the white line position with almost the same magnitude
(\fref{Fe_XAS}).  The complex fine structure of the Fe
\Ledge[2,3] XAS is not compatible with a pure Fe$^{3+}$ valency state,
since we observed only one peak -- similar to  \cite{Kanchana2007prb} --
instead of a double peak structure.
In order to provide a quantitative description of the spectral
features, we took into account both Fe valencies $3+$ and $2+$,
separately, using a primitive unit cell.  The theoretically calculated Fe
\Ledge[2,3] XAS agrees most closely with the experimental data by
using those results in a linear combination of \SI{60}{\percent}
Fe$^{3+}$ and \SI{40}{\percent} Fe$^{2+}$ (\fref{Fe_XAS}),
which is opposite to the calculated proportion found in
Ref.~\cite{Besse2002epl} using the framework of the ligand-field atomic
formalism.

Considering that the correct Fe valency in the SFMO ground state is 3+
as concluded in many theoretical studies, 
the mixed valency state might follow again from lattice
defects as speculated in \cite{Kanchana2007prb}. We investigated the influence of the two types of defects in
the tetragonal supercell (\fref{fig:SFMO_structure}).  For the antisite defects, we observed only
the Fe$^{3+}$ solution.  This can be connected with the high defect
concentration modeled within the supercell.  However, a single vacancy  among twelve oxygen atoms models a more realistic concentration.
Self-consistent calculations in the tetragonal supercell produce the
valency of the Fe ions being equal to 2.9+ and 2.4+ at the Fe1 and Fe2
sites, respectively.  Therefore, the existence of the vacancy shifts
the valency of the nearest Fe ion (Fe2) towards 2+. This valency
change could be also observed below in the CPA calculation of the DOS
including oxygen vacancies (\fref{fig:DOS_compare_PES}).

Indeed, the full explanation of the experimental spectra is only
possible by taking into account these crystal imperfections. The Fe
\Ledge[3] x-ray absorption spectrum for left circularly polarized
light ($\sigma^+$) possesses four major fine structures \textit{a}, \textit{b}, \textit{c}
and \textit{d} (\fref{Fe_XMCD}). We found that the calculations for
the ideal crystal structure with the Fe$^{3+}$ ground state solution
(dashed blue line) provides the x-ray absorption intensity $\sigma^+$
only at the major peaks \textit{b}, \textit{c}, and \textit{d} and do not reproduce the low
energy shoulder (peak \textit{a}) as well as the low energy peak \textit{e} at the
\Ledge[2] edge.  The calculations with the Fe$^{3+}$ solution produce
only one high energy peak structure in the \Ledge[3] $\sigma^-$
spectrum (middle panel). However, the experimental measurements
exhibit a double-peak structure.  The x-ray absorption from the Fe2
atoms with the oxygen vacancy nearby (solid red line) contributes to the low
energy peak of $\sigma^-$ absorption (\fref{Fe_XMCD}).  The
relative intensity of the peaks depends on the relative concentration
of the Fe1 and Fe2 ions in SFMO, in other words, the concentration of
defects such as oxygen vacancies.  It is similar for the oxygen \Kedge
edge. The calculations including an oxygen vacancy are in better
agreement with the experimental measurements in the x-ray absorption
as well as in the XMCD (\fref{O_K}).

The XAS and XMCD spectra at the Mo \Ledge[2,3] and
\Medge[2,3] edges are less sensitive for the crystal defects.  For
both edges the agreement with the experimental measurements
 \cite{Kanchana2007prb} is quite good and independent from
 the concentration of oxygen vacancies
(see SI, figures S4 and S5).

\section{Electronic and Magnetic Properties with Randomly Distributed Oxygen Vacancies}
\label{sec:oxygen_vacancies}
We saw above that oxygen vacancies are certainly present in most samples of
SFMO, while their impact on electronic and magnetic properties should be studied 
in detail. The KKR-GF method combined with the coherent potential approximation (CPA)
allows the study of arbitrary concentrations of randomly distributed
oxygen vacancies as done before for SrCoO$_{3-\delta}$ \cite{Hoffmann2015prb_own}.
The oxygen vacancies were simulated inside the primitive
unit cell with one functional unit of SFMO (see blue arrows in
\fref{fig:SFMO_structure}).
The internal lattice positions had to be kept static \cite{Burzo2011}.  
We introduced a
certain percentage of empty spheres at the lattice
sites of the oxygen ions, which are modeled as randomly distributed via CPA.  The typical
oxygen-deficiency $\delta$ ranges between 0.006 to
0.36 \cite{Kircheisen2012,Yamamoto2000JMaterChem, Colis2005jap_10.1063/1.1997286}. This
represents \SIrange{0.1}{6}{\atper} of the total oxygen amount in
defect-free SFMO.

\begin{figure}[b]
  \includegraphics[width=236.68pt]{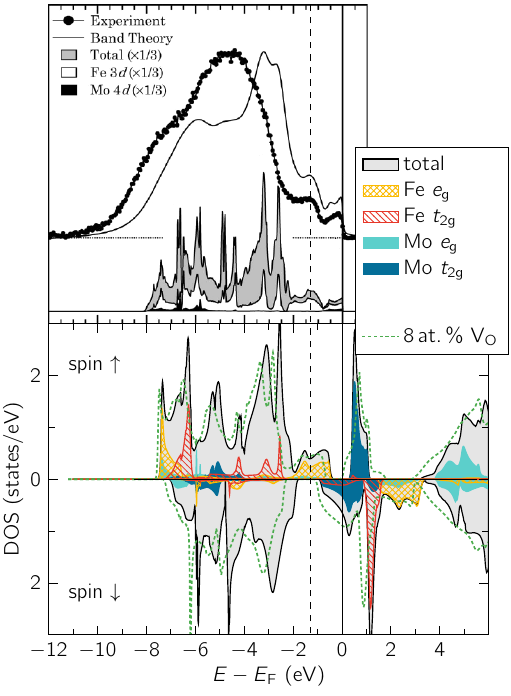}
  \caption{(Color online) 
    The upper panel shows 
    the experimental and simulated PES of SFMO taken 
    as a copy of Fig. 5(a) of Ref. \cite{Saitoh2002PRB} 
    (some labels were removed). The lower 
    panel represents the DOS for SFMO calculated with the 
    KKR-GF method and $\UFe=\sieV{2}$.
    The gray area shows the total DOS for defect-free SFMO.  
    The colored regions
    indicate the LDOS for the $d$ states of Fe (reddish) and Mo
    (bluish).
    The green dotted DOS includes \SI{8}{\atper} of randomly
    distributed oxygen vacancies. Both plots are scaled with respect
    to the same energy scale and the Fermi energy at zero
    (black vertical line). The dashed line indicates
    the position of the Fe $\due$ state in the
    experimental PES.}
  \label{fig:DOS_compare_PES}
\end{figure}

We obtained for defect-free SFMO 
the half-metallic ground state by applying
correlation corrections following Refs.
\cite{Szotek2003prb, Stoeffler2006jpcm, Munoz-Garcia2011}
(\fref{fig:DOS_compare_PES}). 
Although the half-metallic ground state would be well represented within the 
self-interaction correction (SIC) \cite{Szotek2003prb},
the Fe $d$ states localize too much.
Hence, we applied an $U$ parameter on the $d$ states of Fe with $\UFe=\SI{2}{\eV}$
as a compromise,
because we are aware that the correct treatment of the electronic 
structure is very sensitive to the structural \cite{Solovyev2002prb}
and methodological differences
\cite{Kobayashi1998n, Sarma2000PRL,Reyes2016jpcc} (see SI).
Following optical \cite{Tomioka2000} and photoemission
spectroscopy (PES) \cite{Saitoh2002PRB, Kang2002prb,Kim2003jkps} measurements,
the band gap should be \SIrange{0.5}{1.3}{\eV}, and should 
open between the Mo $\dut$ states and the Fe $\due$ states.

Of course, a direct comparison of calculated DOS and experimental spectra
is very complicated, since thermal broadening and defect levels can 
easily reduce the ideal band gap.
Saitoh \etal \cite{Saitoh2002PRB} noticed for example possible ASD
in their samples (up to \SI{10}{\atper}) but 
could not account for defects in their theoretical interpretation.
ASD inside SFMO deteriorate the DOS at the Fermi energy $\EF$
because new states appear in the majority spin channel 
\cite{Solovyev2002prb, Stoeffler2005jpcm, Munoz-Garcia2011, Reyes2016jpcc}
and reduce the spin polarization, which was indeed observed
experimentally \cite{Panguluri2009ApplPhysLett}. 
In contrast, a low concentration of oxygen vacancies does not alter the half-metallic 
character of SFMO \cite{Stoeffler2005jpcm, Munoz-Garcia2011, Wu2014ssc},
but additional Fe $\dd$ and Mo $\dd$ states become occupied  
\cite{Munoz-Garcia2011}. This indicates a lower valency of the Fe
and Mo ions. Electrons of the removed oxygen atoms occupy
states of Fe and Mo ions. A similar qualitative behavior 
was observed within our calculations of oxygen-deficient SFMO with CPA
(green dashed line in \fref{fig:DOS_compare_PES}).
The averaging within the CPA broadens all states. 


\begin{table}
  \caption {The calculated spin $m_\text{s}$ and orbital $m_\text{o}$
    magnetic moments (in $\muB$) of Fe and Mo in defect-free SFMO
    compared with the measured magnetic moments. 
    Experimental uncertainties are given in brackets behind the
    values.}
  \label{tab:mom}
  \lineup
  \begin{indented}
    \item[]
    \begin{tabular*}{\columnwidth}{l|@{}S[table-format=2.2(2)]S[table-format=1.3(2)]
        |S[table-format=2.3(2)]S[table-format=1.3(2)]}
      \br
      & \multicolumn{2}{c|}{Fe} & \multicolumn{2}{c}{Mo}\\
      method                   & $m_\text{s}$             & $m_\text{o}$& $m_\text{s}$             & $m_\text{o}$ \\
      \hline
      SPR-LMTO                 &    4.104           & 0.041 &-0.504           & 0.047 \\
      KKR-GF                   & 3.921  &   & -0.525  &   \\\mr
      Ref.~\cite{Besse2002epl}$^1$     &  3.05 \pm 0.20 &  0.02 \pm 0.02 &-0.32 \pm 0.05 & -0.05 \pm 0.05\\
      Ref.~\cite{Koide2014jpcser}$^1$  & 2.80 \pm 0.30 &    0.093 \pm 0.010 &-0.36 \pm 0.03 & -0.037 \pm 0.015\\    
      Ref.~\cite{Garca-Landa1999}$^2$  & 4.1 \pm 0.1 &  & 0.0 \pm 0.1       &    \\
      Ref.~\cite{Kanchana2007prb}$^3$  & 3.72              & 0.042 &-0.29            & 0.020\\
      Ref.~\cite{Fang2001prb}$^3$      & 3.97              &  &-0.39            &  \\
      \br
    \end{tabular*}
    \item[] $^1$XMCD measurement
    \item[] $^2$Neutron diffraction experiment
    \item[] $^3$Theory
  \end{indented}
\end{table}

Even more interesting is the effect of oxygen vacancies on the 
magnetic properties of SFMO. As a starting point, we compare briefly
the numerically obtained magnetic properties for defect-free SFMO
with measured results and introduce then oxygen vacancies again. 
The expected ferrimagnetic (FiM) ground state of defect-free SFMO
was always energetically favorable when comparing possible
magnetic configurations.
The Fe spin and
orbital moments in FiM are parallel, whereas the spin and orbital Mo moments
are antiparallel to each other, in accordance with Hund's third
rule (\tref{tab:mom}). 
We obtain a good agreement for the orbital Fe
magnetic moment ($\simB{0.041}$) with the FPLMTO result of
Jeng and Guo \cite{Jeng2003prb} ($\simB{0.043}$) and the LMTO results of
Kanchana \etal \cite{Kanchana2007prb} ($\simB{0.042}$). The calculated
spin and orbital moments of the Mo atom are always larger in comparison with earlier
calculations \cite{Saha-Dasgupta2001prb, Kobayashi1998n, Fang2001prb,
 Jeng2003prb, Kanchana2007prb}.  
Also, the Fe spin moment
of $\approx\simB{4}$ is in a good agreement with the experimental
neutron diffraction measurements \cite{Garca-Landa1999}
and the theoretical ideal value for the \Feval{3}/\Moval{5} or
\Feval{2}/\Moval{6} valency configuration (\tref{tab:mom}). 
However, typical values for the magnetization reported in 
experimental studies are smaller than $\approx\simB{4}$.
This reduction was always attributed to point defects
\cite{Ogale1999apl, Balcells2001apl, Besse2002epl, Kircheisen2012}.

\begin{figure}
  \includegraphics[width=236.68pt]{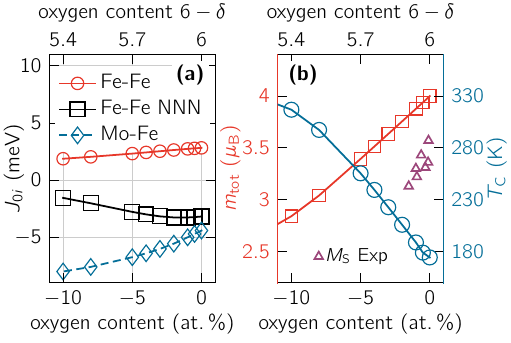}
  \caption{ Calculated magnetic properties depending on the 
    oxygen deficiency $\delta$ obtained with the KKR-GF method.
    (a) Magnetic exchange coupling constants for the strongest 
    interactions (displayed in \fref{fig:SFMO_structure}).
    (b) Total magnetic moment $m_\text{tot}$ obtained within the 
    self-consistent calculations and Curie temperatures from the 
    Monte Carlo simulation.  
    Experimental results \cite{Kircheisen2012} are given 
    for the saturation magnetization (in $\muB$).
  }
  \label{fig:mag_moment_V_O}
\end{figure}

We used the CPA to simulate randomly distributed defects and 
indeed an almost linear decrease of the total magnetic
moment agreeing qualitatively
with an experimental study by Kircheisen \etal \cite{Kircheisen2012}
was observed
(\fref{fig:mag_moment_V_O}(b)). The offset of the measurement 
at $\delta=0$ will result from other point defects like 
ASD \cite{Ogale1999apl}. Surprisingly, the Curie temperature is
increasing with more oxygen vacancies, while the total magnetic moment
is reduced at the same time (\fref{fig:mag_moment_V_O}(b)). We obtained the 
Curie temperature $\Tc$ by calculation of the magnetic exchange
interactions $J_{ij}$
and consider them up to a distance of \SI{12.49}{\angstrom}
in a Monte Carlo simulation with a classical Heisenberg model \cite{Fischer2009prb}.
Thoses result agree also with the mean-field approximation (MFA) or the random-phase 
approximation (RPA) considering that MFA usually overestimates
$\Tc$. 

Although $\Tc$ for defect-free SFMO is below the range of experimental values
\SIrange{324}{420}{\kelvin}, we are interested in the relative variation of $\Tc$
with the amount of oxygen vacancies, which is roughly
$\SI{+15}{\kelvin}$  per \si{\atper} of oxygen vacancies.
Experimental results might therefore be enhanced by 
tens of kelvin due to oxygen vacancies. 

The reasons for this increase in $\Tc$ can be understood from the 
magnetic exchange interactions. 
The most prominent coupling
constants of the order of several \si{\milli\electronvolt} have
only a very restricted range up to \SI{7.9}{\angstrom}
(\fref{fig:mag_moment_V_O}(a)). This
includes only the interactions up to the next nearest neighbor Fe ions
$J^\text{Fe-Fe}_{02}$ (see inset in \fref{fig:SFMO_structure}).
Due to the tetragonal structure, all
magnetic exchange interactions show a small asymmetry with respect to
those with a component in $z$ direction. For clarity, only the
coupling constants in the $x$-$y$-plane are shown
(\fref{fig:mag_moment_V_O}(a)).

The calculations of $J_{ij}$ were performed
in the FiM reference state. We observe for defect-free SFMO a strong
antiferromagnetic (AFM) coupling between Fe and Mo ions and a ferromagnetic
coupling between nearest neighbor Fe--Fe, $J^\text{Fe-Fe}_{01}$. 
The next nearest Fe--Fe coupling constants, $J^\text{Fe-Fe}_{02}$, have also
an AFM character but are not enough to destabilize the FiM ground state of SFMO
(see \fref{fig:mag_moment_V_O}(a) for $\delta=0$).
Introducing now the oxygen vacancies in a random fashion reduces the 
oxygen mediated exchange between the Fe ions but favors the AFM exchange 
between Fe and Mo ions (\fref{fig:mag_moment_V_O}(a)). This behavior resembles a stronger  
orbital localization as observed with an increase of electron correlation 
parameter $\UFe$ (see SI).
The stronger localization of the orbitals leads to a decrease of 
the electron hopping and, thereby, to a decrease in the magnetic coupling strength.  
Together, the total magnetic moment will be indeed reduced but 
the stronger AFM coupling between Fe and Mo sites will at the same time 
mediates an additional FM coupling and increases $\Tc$.  

\section{Conclusions}
\label{sec:conclusion}

We focused our study on the impact of oxygen defects on the
magnetic properties of SFMO and made two crucial findings:

Our simulated x-ray absorption spectra (XAS) and 
x-ray magnetic circular dichroism spectra  (XMCD) show a 
better agreement with experimental spectra
\cite{Linden2000apl, Kapusta2002jmmm, Besse2002epl, Kuepper2004pssa}
than previous density
functional calculations \cite{Kanchana2007prb}.
This indicates that the experimentally observed 
mixed valency of the Fe ion (Fe$^{2+}$ or Fe$^{3+}$)
is mainly caused by oxygen vacancies.

Second, we could explain the experimentally observed reduction of the 
total magnetic moment of SFMO by an increasing amount of oxygen
vacancies following from a variation of the FM and AFM magnetic 
coupling between the Fe and Mo ions. At the same time,
the Curie temperature shows a strong increase of roughly
$\SI{+15}{\kelvin}$ per \si{\atper} of oxygen vacancies.

Since the growing conditions of SFMO samples
could be influenced by the oxygen partial pressure, oxygen vacancies
could be a tool to improve the magnetic properties of SFMO, in particular, 
because the half-metallic band gap is not influenced by oxygen vacancies. 

\ack
This publication was funded by the German Research Foundation
within the Collaborative Research Centre 762 (projects A4 and B1).
The support  of the German Academic Exchange Service
by grant ``Electronic properties of thin SFMO
films'' (grant number 57071667) is greatly acknowledged.
\section*{References}
\bibliography{journals,lib}


\section*{Supplementary material}

# Supporting information for: Variation of magnetic properties of Sr$_2$FeMoO$_6$ due to oxygen vacancies

Martin Hoffmann,[*,†] Victor N. Antonov,[‡,¶] Lev V. Bekenov,[‡,§] Kalevi Kokko,[∥,⊥] Wolfram Hergert,[#] and Arthur Ernst[†,§]

†Institute for Theoretical Physics, Johannes Kepler University Linz, Altenberger Straße 69, 4040 Linz, Austria

‡G. V. Kurdyumov Institute for Metal Physics of the N.A.S. of Ukraine, 36 Vernadsky Street, 03142 Kiev, Ukraine

¶Faculty of Mathematics and Informatics, University of Bialystok, K. Ciolkowskiego 1M, PL-15-245 Bialystok, Poland

§Max Planck Institute of Microstructure Physics, Weinberg 2, 06120 Halle, Germany

∥Department of Physics and Astronomy, University of Turku, FIN-20014 Turku, Finland

⊥Turku University Centre for Materials and Surfaces (MatSurf), Turku, Finland

#Institute of Physics, Martin Luther University Halle-Wittenberg, Von-Seckendorff-Platz 1, 06120 Halle, Germany

E-mail: martin.hoffmann@jku.at

## Simulation of x-ray spectra

The x-ray absorption and dichroism spectra were calculated with the spin-polarized fully relativistic linear-muffin-tin-orbital (SPR-LMTO) method taking into account the exchange

splitting of the core levels. The approach for the x-ray absorption spectroscopy (XAS) and x-ray magnetic circular dichroism (XMCD) simulations is described in our previous papers.[S1] The finite lifetime of a core hole was accounted for by folding the spectra with a Lorentzian. The widths of core level spectra $\Gamma_{L_2} = 1.14\,\text{eV}$ and $\Gamma_{L_3} = 0.41\,\text{eV}$ for Fe, $\Gamma_{L_2} = 1.83\,\text{eV}$, $\Gamma_{L_3} = 1.69\,\text{eV}$ and $\Gamma_{M_{2,3}} = 2.1\,\text{eV}$ for Mo were taken from.[S2] The finite apparative resolution of the spectrometer was accounted for by a Gaussian of width $0.6\,\text{eV}$.

In our SPR-LMTO simulations, the basis set consisted of the $s$, $p$, and $d$ LMTO's for the Sr, Fe, Mo and O sites. The $\boldsymbol{k}$-space integrations were performed with the improved tetrahedron method [S3] and the self-consistent charge density was obtained with 1063 irreducible $\boldsymbol{k}$-points. We used a relativistic generalization of the LDA+$U$ method, which takes into account the spin-orbit coupling so that the occupation matrix of localized electrons becomes non-diagonal in spin indexes. This method is described in detail in our previous paper [S4] including the procedure to calculate the screened Coulomb $U$ and exchange $J$ integrals, as well as the Slater integrals $F^2$, $F^4$, and $F^6$. The constrained LSDA calculations produce $J = 0.85\,\text{eV}$ for the Fe site in SFMO. In order to be consistent with the other methods, the Hubbard $U$ has been used as an external parameter.

Table S1: The calculated spin $m_\text{s}$ and orbital $m_\text{o}$ magnetic moments (in $\mu_\text{B}$) of Fe and Mo in defect-free SFMO compared with the LSDA+$U$ method (SPR-LMTO) and the results obtained via sum rules.

|          | Fe    |       | Mo     |       |
|----------|-------|-------|--------|-------|
| method   | $m_\text{s}$ | $m_\text{o}$ | $m_\text{s}$ | $m_\text{o}$ |
| SPR-LMTO | 4.104 | 0.041 | −0.504 | 0.047 |
| Sum rules | 4.283 | 0.052 | −0.464 | 0.041 |

**Sum Rules**  We note that the spin and orbital magnetic moments were obtained from the XMCD experiments by using the sum rules, which relate the integrated signals over the spin-orbit split core edges of the circular dichroism to the ground state orbital and spin magnetic moments.[S5] It is well known that the application of the sum rules sometimes results in an error up to $50\,\%$.[S6] To investigate the possible error of the sum rules in the case of SFMO we

compare the spin and orbital moments obtained from the theoretically calculated XAS and XMCD spectra through sum rules with directly calculated LSDA+$U$ values (table S1). The number of the transition metal $3d$ electrons is calculated by integrating the occupied $d$ local density of states inside the corresponding atomic spheres, which gives the values $n_d^{\text{Fe}} = 6.737$ and $n_d^{\text{Mo}} = 4.552$. The sum rules reproduce the spin magnetic moments within $4\,\%$ and $8\,\%$ and the orbital moments within $21\,\%$ and $14\,\%$ for Fe and Mo, respectively (table S1).[S7–S9]

**XAS and XMCD for Mo** Since the XAS and XMCD spectra were not influenced by the oxygen vacancies, we are here showing the obtained results for the L and M edge of the Mo ion for the sake of completeness (figures S1 and S2).

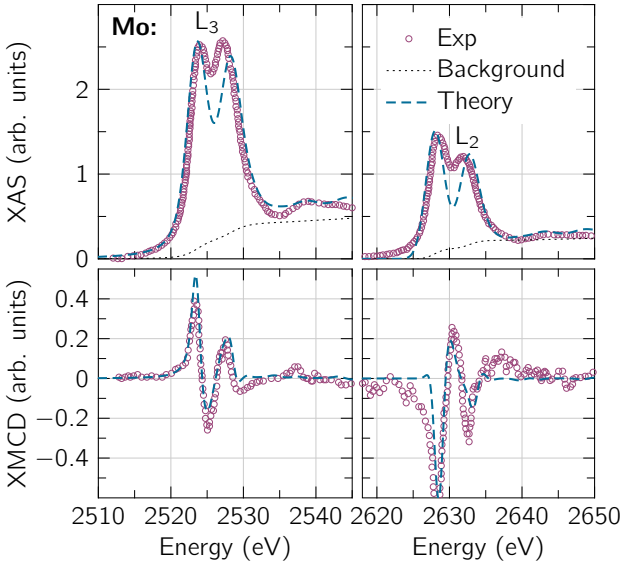

Figure S1: The experimental x-ray absorption spectra (Ref. [S10], open circles) at Mo $L_{2,3}$ edges (top panels) in SFMO measured at $10\,\text{K}$ with $5\,\text{T}$ magnetic field and experimental XMCD spectra (lower panels) compared with the theoretical simulations.

# Electronic Structure

The orbital resolved density of states (DOS) was calculated with the Korringa-Kohn-Rostoker Green's function (KKR-GF) method called HUTSEPOT within the full-charge density approximation to the crystal potential. For the Green's function, the angular momentum cutoff

S3

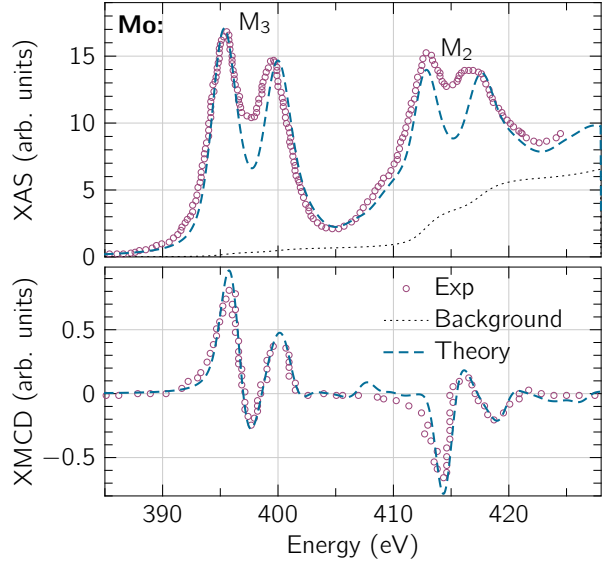

Figure S2: The experimental x-ray absorption spectra (Ref. [S11], open circles) at Mo $M_{2,3}$ edges (top panels) in SFMO and experimental XMCD spectra (lower panels) compared with the theoretical simulations.

was $l_{\max} = 3$ and the Brillouin zone integration was done with a $\boldsymbol{k}$-point mesh of $12 \times 12 \times 12$. The complex energy contour was integrated over 24 Gaussian quadrature points. With the Green's function $G(E)$ of the system, all quantities of interest follow in a straightforward way.

The exchange-correlation functional of a GGA-type was used very successful in the context of SFMO.[S12,S13] Within the KKR-GF method, we used in particular the version of Perdew, Burke and Ernzerhof (PBE).[S14,S15] In order to deal with the strong correlations in the oxide system, the GGA+$U$ approach of Dudarev *et al.* [S16] were used ($U_{\mathrm{eff}} = U - J$).

We note, that the calculation of the electronic structure in SFMO is very sensitive to the choice of the atomic radii and the crystalline structure.[S17] We used the atomic radii, obtained from the largest overlap of atomic potentials. Using smaller radii for the Fe atoms led to a half-metallic solution independent of the exchange-correlation functional – either LDA or GGA. We observed the same also within the SPR-LMTO method.

**Half-metallic SFMO**  Although the half-metallic solution was previously obtained within the LDA and the GGA with different calculation techniques,[S12,S19] the KKR-GF method

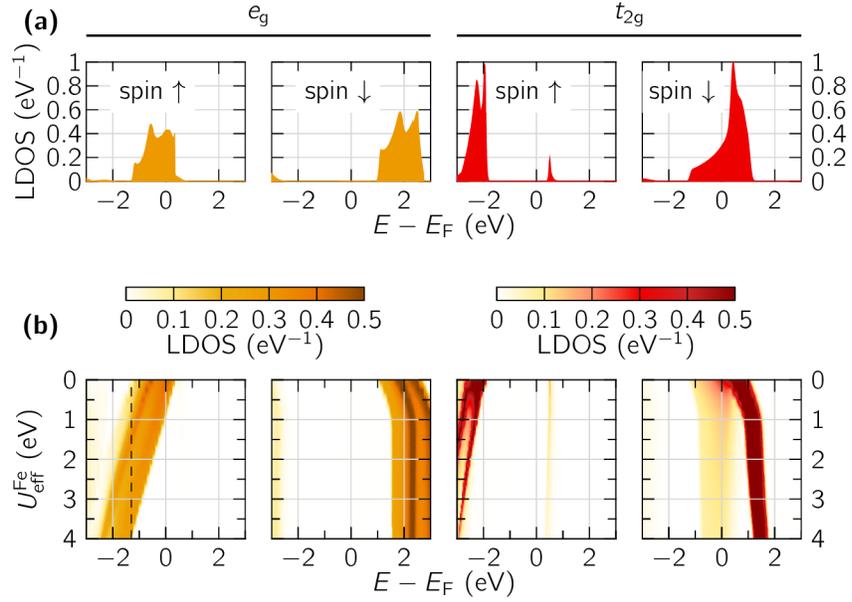

Figure S3: Atomic and orbital resolved DOS for the Fe $d$ states: (a) GGA calculation and (b) GGA+$U$. In (b) the DOS is color-coded as a contour plot. The dashed line in the left hand side panel indicates the peak position in the experimental PES.[S18]

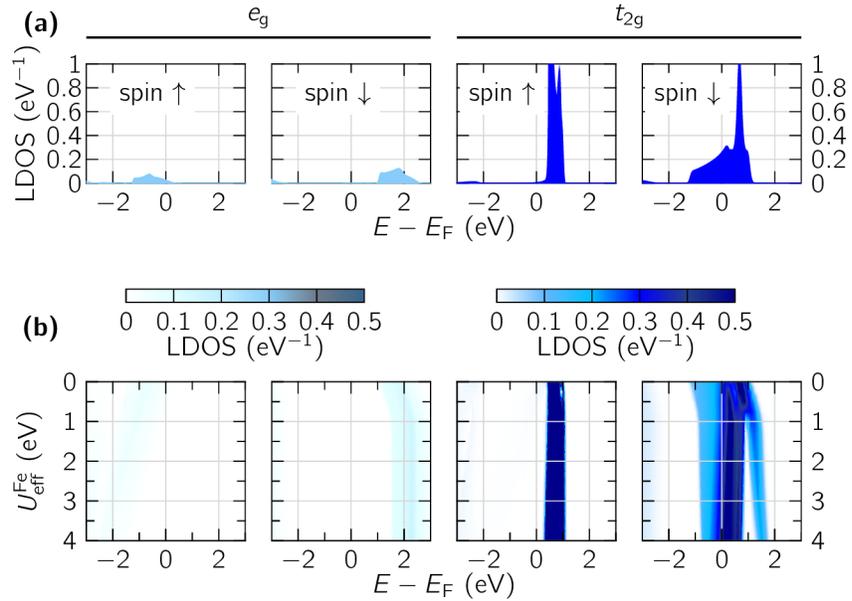

Figure S4: Atomic and orbital resolved DOS for the Mo $d$ states: (a) GGA calculation and (b) GGA+$U$. In (b) the DOS is color-coded as a contour plot.

yielded for the experimental lattice structure of Burzo *et al.* [S20] a metallic solution similar as in references [S21–S23] with a nonzero density of states in both spin channels at the Fermi energy, $E_\text{F}$, independent of the use of LSDA or GGA (for GGA see figures S3a and S4a). We also crosschecked the calculation of the DOS with the other methods and found similar results. The key point here seems to be the lattice structure, because small changes in the Fe-O and Mo-O bond length can influence the existence of the half-metallic band gap.[S17]

In order to obtain a better agreement for the Fe $e_\text{g}^\uparrow$ state with the experimental observations, we calculated the DOS for several correlation parameters $U_\text{eff}^\text{Fe}$ applied only to Fe $d$ states, and found a good agreement for $U_\text{eff}^\text{Fe} = 2\,\text{eV}$ (see dashed line in figure S3b). The main changes occur for the Fe $e_\text{g}^\uparrow$ states at $E_\text{F}$, which crossed $E_\text{F}$ for $U_\text{eff}^\text{Fe} \gtrapprox 1\,\text{eV}$ and the band gap emerged. Then, almost all $d^\uparrow$ ($d^\downarrow$) states of Fe are occupied (unoccupied), which is represented by a formal $Fe^{3+}$ valence state. For the Mo ions, almost all states are unoccupied, except the delocalized $t_{2\text{g}}^\downarrow$ states, which could be considered as a $Mo^{5+}$ state. This situation was found in all self-consistent solutions with GGA+$U$ and represents the generally accepted ground state of SFMO.[S12,S22,S24,S25] Thus, we use it as well in the description of the magnetic properties below.

## Magnetic properties

We calculated the interatomic exchange coupling parameters $J_{ij}$ from *ab initio* using the magnetic force theorem [S26] implemented within HUTSEPOT. The $J_{ij}$ can be used in a Monte Carlo (MC) simulation to obtain the critical temperatures (details in references [S27–S29]). The convergence of the results concerning the number of MC steps, the cluster size, and the number of included $J_{ij}$ was tested. We found that a cluster of $20 \times 20 \times 20$ unit cells is a good choice (figure 1 in the paper). We considered also periodic boundary conditions and restricted the calculation only to the magnetic atoms of the unit cell. The starting point was a high-temperature disordered state above the critical temperature $T_\text{C}$. In the course of

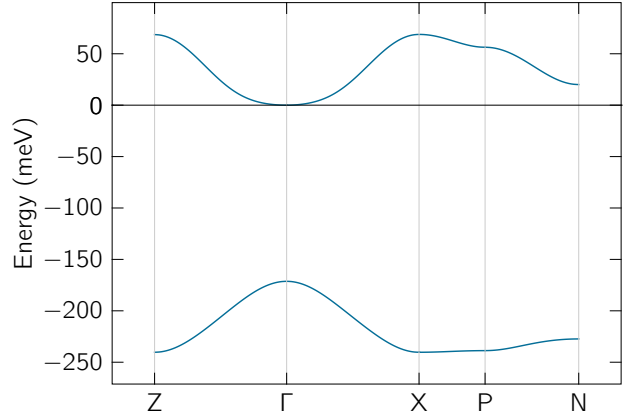

**Figure S5:** Magnon spectrum of half-metallic SFMO ($U_{\text{eff}}^{\text{Fe}} = 2\,\text{eV}$). The positive (negative) mode is associated mainly with the magnetic moments of Fe (Mo).

the simulations, the temperature was stepwise reduced until magnetic ordering was reached. For each temperature $T$, 20 000 MC steps are enough for the thermal equilibrium and the thermal averages were determined over 20 000 additional MC steps. $T_\text{C}$ was then obtained from the temperature dependency of the magnetic susceptibility and the heat capacity within a numerical uncertainty range of $\pm 5\,\text{K}$. For comparison, $T_\text{C}$ was also calculated with the mean-field approximation (MFA) and random-phase approximation (RPA).[S30]

**Magnetic ground state** The underlying magnetic reference state for the calculation of the $J_{ij}$ was the ferrimagnetic (FiM) solution – Fe moments align parallel to each other and Mo moments align antiparallel to the Fe moments. This was also identified as stable magnetic ground state by the calculation of the corresponding magnon spectrum (figure S5). The spectrum shows two well separated modes. In accordance with our analysis of the corresponding site-resolved magnonic Green function, the positive mode is mainly associated with the collective precession of Fe magnetic moments, while the negative mode reflects the properties of Mo magnetic moments. The different signs of the modes indicate on a ferrimagnetic ground state, which is confirmed by our total energy calculations.

However, Solovyev [S17] found that the FiM is not stable with respect to a spin spiral. The difference to our result might follow from our choice of taking the experimental lattice

structure of SFMO instead of the undisturbed cubic lattice structure.[S17] Nevertheless, the FiM ground might be stabilized including breathing distortions of the oxygen octahedra. If the oxygen ion came closer to the Fe ions, the proposed spin spiral ordering became less favorable with respect to the FiM state but the half-metallicity was gone. These observations match ours. Our calculations within the GGA yielded for the experimental structure (where $d_{\text{Fe-O}}/d_{\text{Fe-Mo}} < 0.5$) a metallic ground state as well.[S20]



\end{document}